\documentclass[10pt,conference]{IEEEtran}

\usepackage[sort&compress,numbers]{natbib}
\usepackage[pdftex]{graphicx}
\usepackage{amsmath}
\usepackage{algorithm}
\usepackage{algpseudocode}
\usepackage{array}
\usepackage{multirow}
\usepackage{color,soul}
\usepackage[dvipsnames]{xcolor}
\usepackage{url}

\usepackage[font=footnotesize]{subfig}

\newcolumntype{P}[1]{>{\centering\arraybackslash}p{#1}}
\newcolumntype{L}[1]{>{\arraybackslash}p{#1}}
\newcolumntype{M}[1]{>{\centering\arraybackslash}m{#1}}

\newif\ifnotes
\notestrue 

\newcommand{\frameit}[2]{
	\begin{center}
		{\color{BrickRed}
			\framebox[3.3in][l]{
				\begin{minipage}{3in}
					\inred{#1}: {\sf\color{Black}#2}
				\end{minipage}
			}\\
		}
	\end{center}
}
\ifnotes
\newcommand{\inred}[1]{{\color{BrickRed}\sf\textbf{\textsc{#1}}}}
\newcommand{\note}[1]{\frameit{Note}{#1}}
\newcommand{\todo}[1]{\inred{$<<<${#1}$>>>$}}
\newcommand{\inote}[1]{\inred{$<<<${#1}$>>>$}}
\else
\newcommand{\inred}[1]{#1}
\newcommand{\note}[1]{}
\newcommand{\todo}[1]{}
\newcommand{\inote}[1]{}
\fi

\newcommand{\comment}[1]{}

\begin{document}
\title{ATHAFI: Agile Threat Hunting And Forensic Investigation}








\author{Rami~Puzis, Polina~Zilberman, 
        and Yuval Elovici
\thanks{R. Puzis at email: puzis@bgu.ac.il}
\thanks{P. Zilberman at email: polinaz@bgu.ac.il}
\thanks{Y. Elovici at email: elovici@bgu.ac.il}
\thanks{Telekom Innovation Laboratories, 
	    Ben-Gurion University of the Negev, 
        Beer-Sheva, Israel.}
\thanks{Department of Software and Information Systems Engineering, 
        Ben-Gurion University of the Negev, 
        P.O.B. 653 Beer-Sheva, Israel.}
\thanks{Manuscript received month day, year; revised month day, year.}
}

\maketitle

\begin{abstract}
Attackers rapidly change their attacks to evade detection.
Even the most sophisticated Intrusion Detection Systems that are based on artificial intelligence and advanced data analytic cannot keep pace with the rapid development of new attacks.
When standard detection mechanisms fail or do not provide sufficient forensic information to investigate and mitigate attacks, targeted threat hunting performed by competent personnel is used. 
Unfortunately, many organization do not have enough security analysts to perform threat hunting tasks and today the level of automation of threat hunting is low.

In this paper we describe a framework for agile threat hunting and forensic investigation (ATHAFI), which automates the threat hunting process at multiple levels. 
Adaptive targeted data collection, attack hypotheses generation, hypotheses testing, and continuous threat intelligence feeds allow to perform simple investigations in a fully automated manner.
The increased level of automation will significantly boost the analyst's productivity during investigation of the harshest cases.

Special Workflow Generation module adapts the threat hunting procedures either to the latest Threat Intelligence obtained from external sources (e.g. National CERT) or to the likeliest attack hypotheses generated by the Hypotheses Generation module. 
The combination of Attack Hypotheses Generation and Workflows Generation enables intelligent adjustment of workflows, which react to emerging threats effectively.


\end{abstract}



\section{Introduction}
\label{intro}

Over the past years, the perpetrators of cyber-attacks play a dynamic cat and mouse game with the defenders.
Preventing a particular kind of a cyber-attack does not mean the hackers give up, but merely that they change their attack technique.
Attackers are constantly on the lookout for new victims, new attack vectors, and new exploits. 
Current attack industry has established methodologies of research, development, and execution of new attacks including attacks with the highest level of sophistication, known as ``Advanced Persistent Threats'' (APTs)\footnote{U.S. National Institute of Standards and Technology (NIST), \url{https://csrc.nist.gov/glossary/term/advanced-persistent-threat}}. 
\comment{
The U.S. National Institute of Standards and Technology (NIST), defines APT as ``an adversary that possesses sophisticated levels of expertise and significant resources which allow it to create opportunities to achieve its objectives by using multiple attack vectors (e.g., cyber, physical, and deception). 
These objectives typically include establishing/extending footholds within the information technology infrastructure of the targeted organizations for purposes of information exfiltration, undermining or impeding critical aspects of a mission, program, or organization; or positioning itself to carry out these objectives in the future. 
The advanced persistent threat: 
(i) pursues its objectives repeatedly over an extended period of time; 
(ii) adapts to defenders' efforts to resist it; and 
(iii) is determined to maintain the level of interaction needed to execute its objectives.''
}

The most effective APT mitigation operations today include routine investigations by security operation center (SOC) personnel, counter intelligence activities by national Computer Emergency Response Teams (CERT), and threat hunting initiated due to security alerts or external threat intelligence.

Threat hunting is one of the most important security operations which includes active collection of forensic evidence in order to discover advanced attacks that are evading existing security solutions~\cite{Haber2018}.
Current state of the art distinguishes between two types of threat hunting: reactive~\cite{thomas2017dynamic}\footnote{\url{https://www.demisto.com/},\url{https://www.mcafee.com/enterprise/en-us/products/investigator.html}} and proactive~\cite{Sqrrl,tylertech}.
Proactive threat hunting relies on cyber threat intelligence (CTI) \cite{tylertech,Samtani2017,rasheed2017threat} in order to formulate attack hypotheses and actively search for potentially malicious behavior~\cite{Lee2018}.
Reactive threat hunting involves forensic investigation and attack hypothesis testing in response to alerts indicating potentially malicious behavior~\cite{Sqrrl}. 
Note that, in both cases, all the information stored in the SIEM can be used in order to better tailor the generated hypotheses to ongoing attacks.
According to SANS~\cite{Lee2018}, 37.3\% respondents who perform threat hunting do so reactively, in response to suspicious events. 
60\% of respondents perform proactive threat hunting either continuously (43.2\%) or at regular intervals (16.7\%). 

Threat hunting is performed by the most skillful security analysts. The number of incidents that the security analysts need to address is very high and there is a huge shortage of security analysts in many countries and there is an urgent need to develop new tools to improve the productivity to the existing ones~\cite{skills-gap}.


Automation during threat hunting helps reducing the time and effort of the analyst, 
increasing the scale and efficiency of hunts across the enterprise, 
and reducing the required analyst qualification for performing certain types of investigations~\cite{Lee2018}.
SIEM analytics, log file analysis, intrusion detection, and alike are largely automated in all current environments.  
The next level of automation is provided by Endpoint Detection and Response (EDR) and Security Orchestration, Automation and Response (SOAR) systems. 
EDR systems trigger adjustments of policies, firewall rules, quarantine, etc., upon predefined alerts.
SOAR playbooks elaborate the response to alerts through prescribed  workflows that may include investigative steps, response, and some of the hunting logic. 
Current playbooks are manually prescribed by analysts based on their expert knowledge and cyber threat intelligence (CTI) reports~\cite{CLAY20155}.

Lee\&Lee~\cite{Lee2018} note that threat hunting cannot be fully automated because threats are moving targets. 
Indeed, a hunting playbook coded a few months ago may become irrelevant when hunting for the newest threats.      

In this paper we present a framework for agile threat hunting and forensic investigation (ATHAFI) addressing the dynamic ever-changing threat landscape and the need for continuous endless forensic investigation which adapts to the current state of the organization and state of the art threats. 
ATHAFI automatically finds support for vague clues contained in collected sensor data based on descriptions of past attacks, which we will further refer to as Indicators of Attacks (IoAs)~\cite{decianno2014indicators}. 
To accomplish this objective ATHAFI reconfigures the sensors to collect data in a targeted manner. 
The proposed framework introduces two innovative components that increase the level of automation during forensic investigations and aid the security officers during strategic decision making. 
The Attack Hypotheses Generation module provides a set of tools, malware, and attack patterns that are used by the attacker with the highest probability during the investigated incident. 
The Workflow Generation module provides the security officers with ready-to-execute distributed threat hunting programs (termed ``workflows'').


This research advances state-of-the-art threat hunting in the following manner: 
\begin{itemize}
	\item ATHAFI framework automates large parts of the entire threat hunting loop. 
	\item ATHAFI framework is the first one to combine diverse functionality such as analysis of high level CTI, hypotheses generation, attack hypotheses ranking, workflow generation, and targeted data collection into one unified semi-automated threat hunting process. 
	\item ATHAFI defines a set of requirements and APIs for the must-have functionality allowing flexible implementation of the hunting loop. 
	\item We provide an implementation show case which integrating multiple COTS products into the unified ATHAFI framework.
\end{itemize} 

The rest of the paper is structured as follows.
First, background on CTI is provided in Section~\ref{sec:cti}.
Followed by related work on threat hunting and threat hunting automation in Section~\ref{sec:related}.
Section~\ref{sec:proposed-framework} describes the components and processes of the proposed ATHAFI framework, and Section~\ref{sec:implementation} describes an example implementation of ATHAFI's functional shell.
Finally, Section~\ref{sec:conc} summarizes the paper.  

\section{Background: Cyber Threat Intelligence}
\label{sec:cti}

Cyber Threat Intelligence (CTI) is structured actionable information for identifying adversaries, their motives, goals, capabilities, resources, and tactics. 
CTI includes evidence-based knowledge in the form of measurable events and the context for their interpretation.
CTI increases the ability of the analyst to recognize relevant threats and respond to them in a timely manner~\cite{US-CTI-Sharing,tylertech}.
CTI is a powerful mean to increase efficiency of various security solutions, such as intrusion detection, response, real time analytics, forensic investigation, and threat hunting.

Since no organization possesses a complete understanding of the threat landscape, the importance of CTI lies with the ability to share threat information among partners in a machine-to-machine manner. 
By sharing the who, what, where, how, and when of malicious activities, targeted organizations obtain a holistic view of the threat landscape, thus increasing their cyber security readiness~\cite{US-CTI-Sharing}. 

According to a survey among various cyber security and IT management roles presented by Shackleford~\cite{shackleford2015}, 48\% of the respondents say their use of CTI has reduced incidents through early prevention, and 51\% said they are able to respond more quickly to incidents.
Methods for CTI analysis can be applied to provide SOC analysts with a list of related information, supporting them in the decision making process while handling cyber incidents~\cite{Settanni2017}.

In an effort to formalize a standard language for sharing CTI, DHS Office of Cybersecurity and Communications funded MITRE to develop the Structured Threat Information eXpression (STIX) language\footnote{\url{https://www.mitre.org/capabilities/cybersecurity/overview/cybersecurity-blog/stix-20-finish-line}}.
STIX covers the entire range of cyber security concepts, including observables, indicators of compromise (IOC), attack patterns, tools, malware, threat actors, course of action, and other.
A STIX element is denoted as STIX Domain Object (SDO).
SDOs such as observable and IOCs are considered low-level SDOs, while SDOs such as attack patterns, tools and threat actors are considered high-level SDOs.
Additional CTI languages include Malware Information Sharing Platform (MISP)\footnote{https://www.misp-project.org/}, as well as proprietary languages and ontologies developed by McAfee~\cite{CTI-McAfee} and IntelGraph by Accenture~\cite{CTI-Splunk}.

CTI can be acquired by a victim organization who records the attack investigation artifacts and shares them with peers.
Wheelus et al.~\cite{Wheelus2016} propose a tiered big data architecture for the automation of capturing and handling of network traffic.
They generate features and artifacts that would be promptly available for machine learning algorithms and anomaly detectors.  
Samtani et al.~\cite{Samtani2017} suggest collecting CTI proactively from large international under ground hacker communities without waiting for attacks to happen.
They develop a framework for storing and analyzing malicious assets such as crypters, keyloggers, web, and database exploits collected from the dark web.


\section{Related Work}
\label{sec:related}

Current paper proposes an automated threat hunting framework based on CTI.
Threat hunting is a domain actively developed nowadays by cyber security industry, but receives relatively little attention from academic research.
Generally speaking, threat hunting includes series of active investigative steps that help confirming or refuting attack hypotheses.
This process may include forensic investigations and various analytics whose objective is inferring the attack steps and collecting artifacts.

\subsection{Threat Hunting}
\begin{figure}[!ht]
	\centering
	\includegraphics[width=0.45\textwidth]{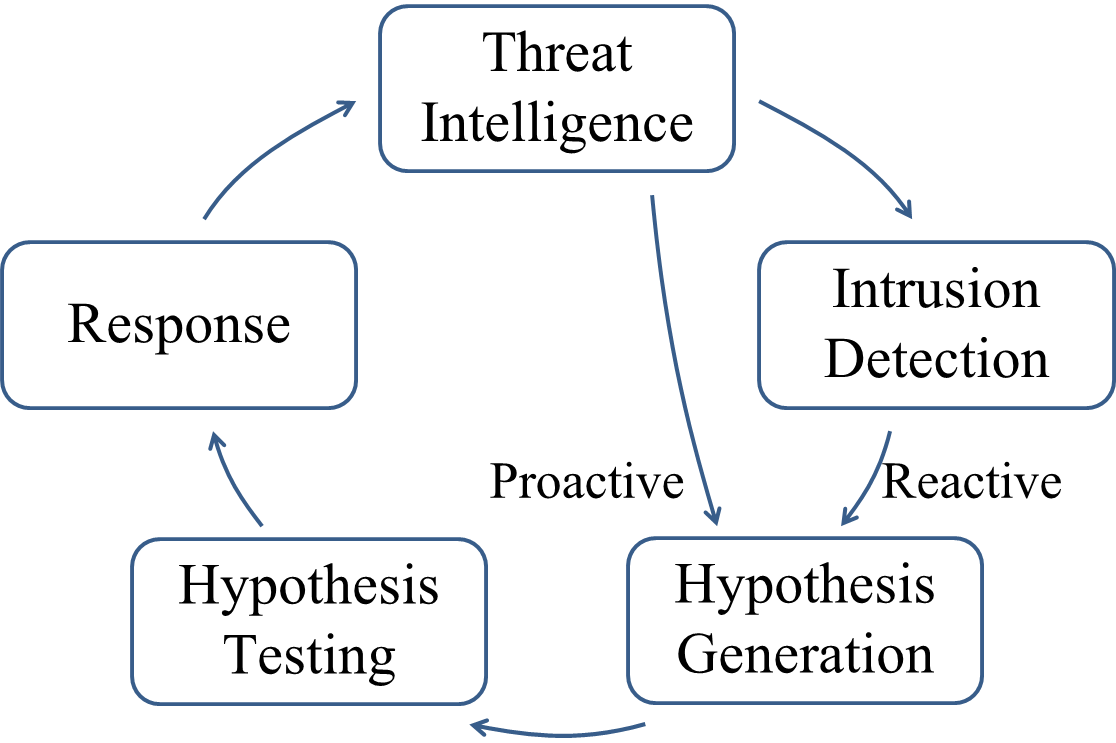} 
	\caption{\label{fig:threat-hunt-life-cycle} Threat Hunting Cycle}
\end{figure}

Cyber security experts are divided on the exact stages of the threat hunting cycle and on its \emph{re-}active or \emph{pro-}active nature.
On one hand, some experts define threat hunting as proactively looking for early indications of presumably ongoing attacks without waiting for alerts to indicate suspicious activity~\cite{Alonso2016}.
On the other hand, threat hunting may refer to an investigative process initiated in response to an alert.
This process may include advanced analytics, forensic investigations, targeted data collection, or policy updating~\cite{rasheed2017threat,Sqrrl}.
The main difference between the proactive and reactive threat hunting is the trigger for the investigation.
Proactive threat hunting relies on \emph{threat intelligence} to actively search for \emph{potentially} malicious behavior.
Reactive threat hunting involves forensic investigation and attack hypothesis testing \emph{in response to} alerts indicating potentially malicious behavior. 
Figure~\ref{fig:threat-hunt-life-cycle} presents the threat hunting cycle comprising both proactive and reactive processes.


Mavroeidis and J{\o}sang~\cite{mavroeidis2018data} present an ontological approach for threat hunting using Sysmon logs in an automated manner.
The authors state the potential benefits of CTI in investigation of attack events and anticipation of the next attack steps, however they do not present hypotheses generation, hypotheses testing, and CTI generation.
Similar to other works they rely on predetermined data collection which is not affected by the course of investigation.

The hypotheses testing phase is elaborated in \cite{Shu2018}, where it is managed by the analyst.
In order to increase the efficiency of hypothesis testing, the authors propose $\tau$-calculus to query CTI and sensor data.
Similar to most current cyber security automation frameworks Shu et al.~\cite{Shu2018} focus on analysis of the data contained in a central storage rather than performing targeted data collection.
 
Thomas et al.~\cite{thomas2017dynamic} also focus on hypotheses testing, but unlike Shu et al.~\cite{Shu2018} they allow the analyst to perform targeted data collection.
The investigation process is guided by the analyst step by step relying on programmable software agents.

Security Orchestration, Automation and Response (SOAR) systems go even further in automation of hypotheses testing.
These kind of systems perform autonomous response to security alert or suspicious indicators relying on prescribed workflows. 
A notable solution that allows targeted data collection with high degree of automation was developed by Demisto Inc.~\cite{demisto-th}.
This product supports automation of threat hunting and incident response by integrating disparate security tools and maintaining a swarm of programmable software agents.
Prescribed investigation workflows, termed \emph{playbooks} in Demisto documentation, encode the steps of forensic investigations triggered by security events such as suspicious or anomalous behavior detected by an IDS.

While Demisto playbooks automate targeted data collection for hypotheses testing, the playbooks themselves are prescribed manually, according to best practices, and are not derived from attack hypotheses or up-to-date CTI.
Moreover, Demisto assume that sophisticated investigations will be manually orchestrated by an analyst because it is prohibitively hard to automatically anticipate the attacker actions in a general manner (generic algorithm that predicts attacker steps).
In contrast, ATHAFI utilizes specific CTI reports to automatically generate investigation workflows which include investigative steps suggested by the report.
To partially mitigate the absence of automated workflows generation Demisto implemented decision support system where the analyst create custom workflows relying on CTI and IntelliSense suggestions derived from existing workflows.

A lot of effort is invested into seamless integration between machine and human analyst~\cite{mcAfee,Lee2018,Sqrrl,tylertech}.
A notable product that provides human-machine teaming capabilities is the McAfee Investigator~\cite{mcAfee}.
This product can be regarded as reactive threat hunting because it starts with choosing an incident for detailed investigation. 
Advanced machine learning algorithms choose the most relevant insights for the human analysts who can then determine the risk and urgency of the incident.
After the analyst has chosen an incident for detailed investigation the machinery uses human input to gather relevant information and reports the summary back to the analyst.

\subsection{Threat Hunting Automation}

Automated threat hunting should not be confused with automated response, e.g. Endpoint Detection and Response (EDR), which includes adjustments of policies, firewall rules, quarantine, blacklisting, etc.
Automated response rules are pre-defined by the expert users and may be provided as the course of action in CTI. 
In hunting, analysts come with new attack hypotheses that need to be confirmed or refuted. 

Experts agree that threat hunting cannot be fully automated~\cite{Lee2018,mcAfee}.
Intuition, creativity, and strategic thinking applied by human analyst are essential to successful threat hunting.
In a SOC context, human intuition is required to find new attack techniques, creativity helps investigating the suspicious cases using available tools, and strategic thinking helps the analyst to scope and assess the security events and make accurate triage decisions. 

Yet, diverse tools can be utilized to automate large parts of the investigative process. 
First, investigations that were already performed and scripted by the analyst can be automated to save future time in performing repetitive, time-consuming tasks~\cite{Sqrrl,Cole2016,microsoftAutoInvest}. 
Demisto Inc.~\cite{demisto} provide their clients with the ability to write playbooks (a.k.a workflows) that encode the steps of forensic investigations.
Furthermore, recently Demisto and Siemplify~\cite{Siemplify} introduced intellisense features, into their playbook writing kit, that recommend arguments and parameters for various investigative tasks based on recordings of past investigations.   
Having the pre-defined playbooks, analysts manually trigger one or more playbooks based on their understanding of current security situation and suggestions presented by the systems.

There is a lack of mechanisms for automatic triggering of workflow executions.
The main reason for this gap is that workflows may be resource intensive and their association with specific alerts may not be trivial.
To partially mitigate this gap threat hunting automation includes advanced analytics~\cite{saikia2019manadac,granadillo2016new}, data visualization in an actionable form~\cite{mcAfee,demisto-th}, human machine teaming~\cite{mcAfee,zhong2018learning}, etc. 

After the analyst has chosen an incident for detailed investigation the machinery uses human input to gather relevant information and reports the summary back to the analyst~\cite{thomas2017dynamic,mcAfee}.
Common threat hunting dashboards strive to present the analyst high level insights and attack hypotheses rather than alerts and IOCs in order to better manage the analyst's cognitive load~\cite{mcAfee, crowdstrike}.
When presented the right information at the right time an analyst can make accurate triage decisions faster and deep dive into the most significant threats.

The most sophisticated solutions utilize interactive guidebooks to help the analysts focus on what is important as they scope and assess the attack hypotheses.
The next generation of threat hunting automation will include fully automated investigations of the simplest attack hypotheses. 
Current state of the art response automation may be turned into a primitive variant of such automatic investigations.
For example, Thomas et al.~\cite{thomas2017dynamic} disclose a system where the first investigative action can be triggered automatically by an alert.
These automated investigations may be significantly expanded by utilizing CTI.  
Although, automated hunting of the most sophisticated threats is too futuristic, the framework proposed in this paper fully automates simple investigations allowing the analyst to focus on the most sophisticated cases. 




\section{The ATHAFI Framework}
\label{sec:proposed-framework}

In this section we describe the main components and processes of the proposed Agile Threat Hunting and Forensic Investigation (ATHAFI) framework as depicted in Figure~\ref{fig:framework}.

\begin{figure}[h]
    \centering
    \includegraphics[width=\columnwidth]{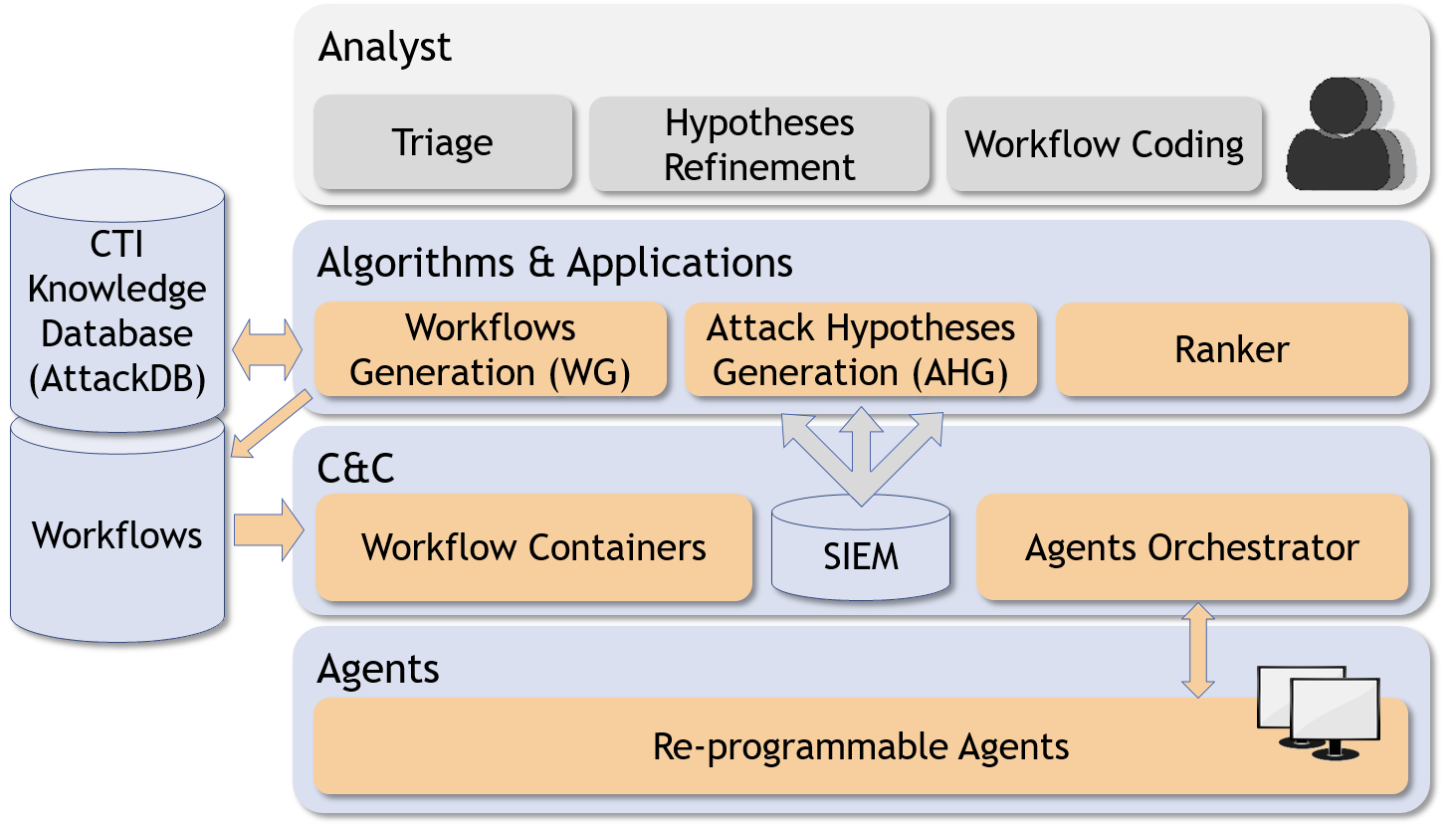} 
	\caption{ATHAFI high-level architecture\label{fig:framework}}
\end{figure}

ATHAFI operates a network of \textbf{re-programmable agents}, able to adapt their behavior to the emerging threats. 
The behavior of the agents is controlled by ready-to-execute distributed \textbf{threat hunting programs (termed ``workflows'')}.
The purpose of these workflows is targeted collection of forensic artifacts in order to confirm or refute an attack hypothesis and track down the attack-path.

The \textbf{command and control unit (C\&C)}, operated by the security analysts, is responsible for managing the workflows. 
In order to execute a workflow, the C\&C pulls it from the workflows database and starts a \textbf{workflow execution container} for each executed workflow.
The C\&C is responsible for: storing predefined workflows, maintaining metadata about workflows in progress, receiving notifications from the workflow containers about workflows' completion of investigation and the existence of forensic findings in the \textbf{Security Information and Event Management (SIEM)} subsystem.

ATHAFI supports both workflows hand-crafted by a security analyst and workflows automatically generated by the \textbf{Workflows Generation (WG)} module. 
In order to generate the workflows, WG relies on Indicators of Attacks (IoAs) provided either from external CTI feeds or generated by the \textbf{Attack Hypotheses Generation (AHG)} module. 
All IoAs are stored in the \textbf{CTI Knowledge Database (AttackDB)}. 
Note, that some of the IoAs stored in the AttackDB are temporary attack hypotheses, i.e. IoAs of presumably ongoing attacks, relevant only during the course of an investigation.
Since it is not feasible to execute workflows for every IoA in the AttackDB, a \textbf{Ranker} module is used to narrow down the searching space for the WG module.
The Ranker also produces a list of IoAs that are most likely to be taking place according to the forensic findings stored in the SIEM.



ATHAFI framework facilitates the entire threat hunting loop, as depicted in Figure~\ref{fig:threat-hunt-life-cycle}.
Re-programmable agents, AHG, Ranker, workflows and WG are the functional core of ATHAFI centered around the AttackDB which stores IoAs provided by CTI feeds. 
CTI stored in AttackDB is used for attack hypotheses generation by the AHG module.
The AHG module allows advanced analysis and correlation between high-level and low-level SDOs.
Sound hypotheses are fed back into AttackDB for further validation by the hypotheses testing workflows. 
Hand-crafted workflows or workflows generated automatically by the WG module test whether or not the organization suffers from one of the attacks described in the AttackDB. 
Response is also facilitated by workflows that encode the required course of action.
ATHAFI orchestrates commercial of the shelf (COTS) systems for intrusion detection and response. 

Flexible design of ATHAFI allows both proactive and reactive threat hunting.  
In the reactive modus operandi, detection of an anomaly 
is followed by generating actionable attack hypotheses and acting to confirm or refute them using the AHG and WG modules, respectively. 
In the proactive modus operandi, both AHG and WG modules act without being triggered by an external event. 

ATHAFI framework facilitates three levels of automation.
\begin{enumerate} 
\item \emph{Reasoning -- } The first level of automation is facilitated by the AHG module where the descriptions of hypothetical attacks (i.e. IoAs) are generated based on collected evidence and CTI. 
\item \emph{Targeted evidence acquisition -- } ATHAFI's workflows automate the forensic investigation process including targeted data collection.
This is similar to the automation level provided by common Security Orchestration, Automation and Response (SOAR) systems (e.g. Demisto's playbooks). 
\item \emph{Closing the loop -- } The last level of automation is facilitated by the WG module that can create workflows from IoAs.     
\end{enumerate}

\subsection{AttackDB Schema}
AttackDB is a central ATHAFI component around which all other modules are revolving. 
The primary objective of AttackDB is encapsulating the knowledge from various CTI sources. 
This knowledge is used by all other modules to perform their tasks. 

AttackDB contains SDOs at all levels of the pyramid of pain (PoP)~\cite{bianco2013pyramid} from the abstract concepts such as tactics and techniques at the top levels down to IOCs and specific observables such as hashes, IPs, and domain names. 
Reasoning on the higher levels of the PoP is challenging and is facilitated by the AHG module. 

\begin{figure}
    \centering
    \includegraphics[width=\columnwidth]{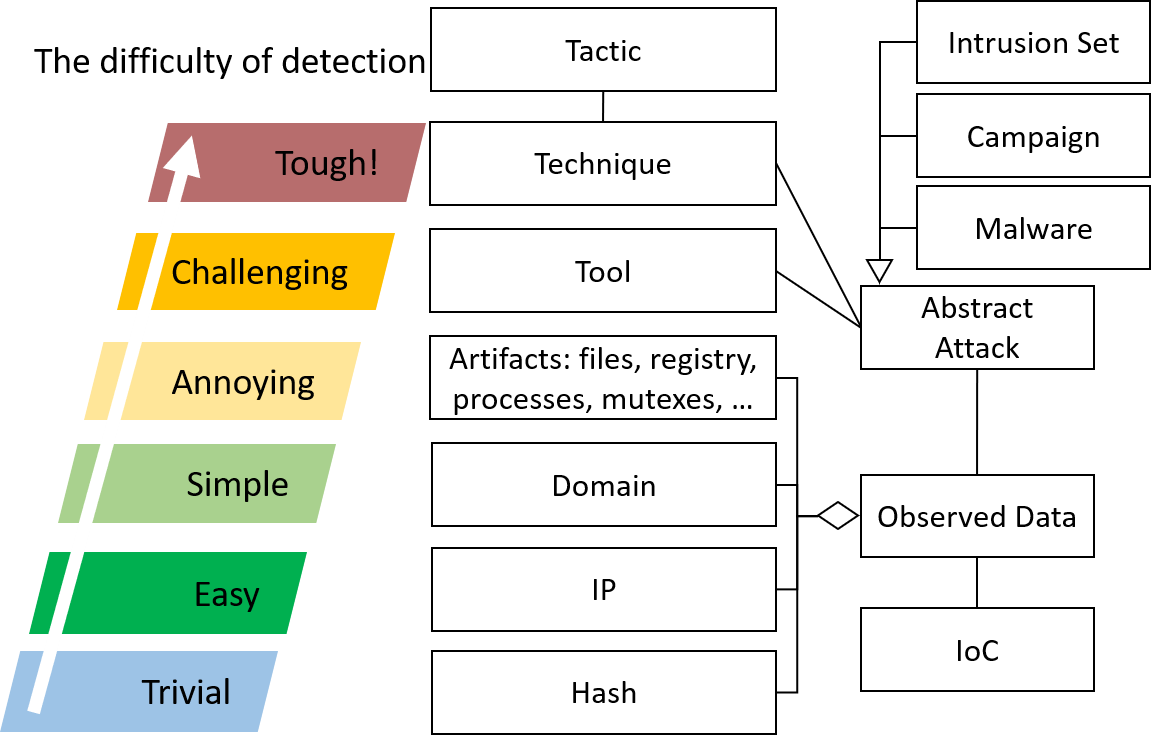}
    \caption{\label{fig:attackdb_pop}AttackDB schema aligned along the Pyramid of Pain}
\end{figure}

Figure~\ref{fig:attackdb_pop} depicts the schematic structure of AttackDB.
The top level SDOs in AttackDB are attack patterns (a.k.a. tactics and techniques). 
These SDOs stand for the malicious activities exhibited by Malware, Campaign or Intrusion Set.  
Malware is a software that exhibits a set of malicious activities. 
Malware can be a part of multiple Campaigns. 
Campaign represents a set of malicious activities at a specific period of time against specific targets. 
Campaigns that are believed to be orchestrated by the same Threat Actor may be grouped into Intrusion Sets. 
Despite the semantic differences between them, Malware, Campaign, and Intrusion Set SDOs can be used to represent an attack being hunted down.

Further AttackDB contains observed data associated with each attack as reported by then CTI. 
Observed data contains hashes, IP addresses, domains, and network and host artifacts (i.e., telemetry) such as process names, services, registry keys, etc. 
Observed data (usually hashes of malicious files or IP addresses) may be grouped together in a pattern and tagged as an IOC. 
IOCs can be used to identify attacks observed in the past but are usually easily modified by the attacker. 
All the observed data stored in AttackDB is processed and used to create attack hypotheses, rank them, and generate workflows.   


\subsection{Agents}

ATHAFI includes software agents installed on PCs and servers across the organization. 
As a general design consideration we prefer using COTS tools developed by market leaders rather than developing tailor-made software in-house for every function which is not the innovation core of the proposed framework.
Specifically, in this research we rely on existing well known (software) agents' infrastructures provided by McAfee ePolicy Orchestrator (McAfee ePO). 


The agents are instructed by workflows to collect data (i.e. observables) which is relevant to selected attacks (or attack hypotheses) according to the AttackDB.    
Collected data is partially processed by the agents in place and the results are transferred to a central repository for deeper analysis. 
Common SIEM products which provide big-data management capabilities as well as a set of correlation, aggregation, and analysis rules can be used. 
The result of the analysis of collected forensic information includes alerts on suspected attacks.

\subsection{Attack Hypotheses Generator}
\label{sec:h-gen}



The Attack Hypotheses Generator (AHG) generates IoAs of hypothetical attacks (hypotheses) based on known attacks given in the AttackDB and the current state of the system, as represented by forensic evidence collected by the agents and stored in the SIEM. 
We refer to an attack hypothesis as a set of SDOs that together describe one complete attack scenario (i.e. Campaign).
While reasoning about attack hypotheses we take into account low level data such as observables and IOCs as well as high level information contained in CTI such as campaigns, tools, malware, attack patterns, etc. 
In the rest of this paper we will refer to tools, malware, and attack patterns collectively as high-level SDOs. 

\subsubsection{Sighting of high-level SDO}
The STIX term sighting denotes the belief that something related to an attack was seen. 
Usually only observables and IOCs, such as registry keys and file hashes, are sighted. 
In some cases tools and malware are sighted as well, when the analyst finds the respective software in the compromised system. 
In addition, observables collected by various sensors and IDSs may lead the analyst to a conclusion that a certain tool, malware, or attack pattern was used by the attacker.
In the following discussions we will use the term sighting to indicate that sufficient evidence was found that a tool, a malware, or an attack pattern was used by the attacker.

\subsubsection{What a good hypothesis is?}
A good attack hypothesis should describe a complete attack scenario starting from the initial penetration throughout the cyber kill chain up until the current state of the attack or even the final goal.  

\begin{figure*}
\centering
\includegraphics[width=\textwidth]{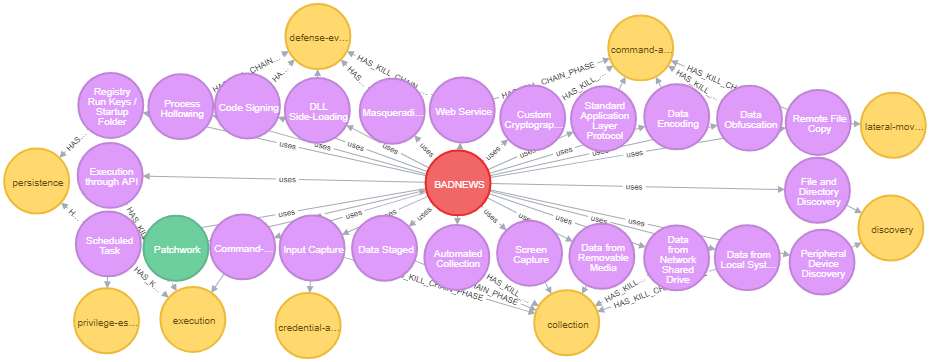}
\caption{BADNEWS malware (red) with its direct relationships to techniques / attack patterns (purple) and an intrusion-set (green). Techniques are connected to tactics (yellow). The set of depicted techniques are considered an IoA for BADNEWS.}

\label{fig:malware}
\end{figure*}

\paragraph{Represent a viable attack}
A good hypothesis should represent a complete attack sequence including if possible initial penetration, lateral movement, persistence, command and control, information collection activities, etc. (see BADNEWS IoA example in Figure~\ref{fig:malware}). 
It is important to include coherent SDOs within a hypotheses such as activities most of which are common to the same threat actor.
An incident report with a detailed description of an attack campaign may serve as an attack hypothesis. 
In fact attacks witnessed by other organizations with similar crown jewels 
are used as attack hypotheses in proactive threat hunting~\cite{Lee2016}. 

\paragraph{Contain actionable insights}
The most important actionable part of a hypothesis is the set of low-level SDOs (IOCs and observables) and high-level SDOs (tools, malware, and attack patterns) that were \emph{not sighted yet}. 
The hypothetic tools, malware, and attack patterns that the threat actor might have been using as a part of the investigated campaign 
should be linked to relevant forensic information in order to    
provide the analyst with directions for further investigation and facilitate automated testing later on.
It is important to include diverse observables which can be partially matched and not only file hashes. 

\paragraph{Being supported by data}
Assume two different attack hypotheses, one presumably attributed to Group A and the other one to Group B. 
The former hypothesis has some support within the data stored in SIEM while the latter does not. 
Having all other parameters equal, which hypothesis should be investigated first?  
Probably the one with some support.

\subsubsection{Hypothesis generation}
The input to the AHG are the IOCs identified so far and stored in the SIEM and a knowledge base composed from structured descriptions of past campaigns. 
AHG algorithms infer the possible high-level SDOs related to the collected IOCs. 
Then, AHG performs higher level reasoning about yet unnoticed tools, malware, and attack patterns and their related IOCs. 
Finally, AHG outputs sets of SDOs (including IOCs) as attack hypotheses.  

Attack hypotheses generation may be modeled for example as a plan recognition problem or a recommendation system. 
In the former case, the AttackDB is the complete attack-graph~\cite{ou2005mulval} of an organization with exploits and attack patterns modeled as actions with pre- and post-conditions.
Given partial observations of the actions the solver reconstructs the full attack path to either one of the possible attack goals~\cite{mirsky2019new}. 
In the latter case, the AttackDB contains a cyber ontology, such as ATT\&CK~\cite{strom2017finding}, linking past incidents to low- and high-level SDOs. 
Relying on this ontology, recommender algorithms produce an ordered list of SDOs that are affiliated with a given set of observed SDOs~\cite{TODO-Aviad}.    

Following the automated generation of attack hypotheses it is also important to allow the security analyst to inspect and augment the hypotheses. 
An automatically generated hypothesis is the first level of automation (i.e \emph{Reasoning}) supporting the decision making of the analyst and inspiring new investigation directions.   
However, the experience, creativity, and strategic thinking of a human analyst are important for successful threat hunting.

\comment{
\begin{figure}
\centering
\includegraphics[width=1.0\columnwidth]{attack_description.jpg}
\caption{An analyst's hypothesis $AD^{analyst}_{BADNEWS}$ for BADNEWS.}
\label{fig:attack_description}
\end{figure}}

\subsection{Ranker}
\label{sec:ranker}
\comment{\note{Which state of the art techniques infer intrusion sets (IoA) from security alerts (IoC?)? check mitre evaluation of COTs (their ability to detect techniques.)
``ChainSmith: Automatically Learning the Semantics of Malicious Campaigns by Mining Threat Intelligence Reports''
\url{https://ieeexplore.ieee.org/abstract/document/8406617}

I didn't remember a paper on inferring intrusion sets from security alerts (I think because there is a large gap between alerts content to intrusion sets entities.
maybe this paper is relevant: real-time APT detection through correlation of suspicious information flows~\cite{milajerdi2018holmes}

IoA from IOC is more common. The simple ways talk about graph representation, and show connected IoAs to IOCs which were found (most of the time used when analyst search in the database during his investigation).
Section B in the related work is relevant}}

The hypotheses generated by the AHG module need to be prioritized.
The Ranker helps in digital triaging by prioritizing hypotheses according to their likelihood given the information available in the SIEM.  

A na\"ive implementation of the Ranker module can rely on the Jaccard distance between the IOCs listed in the hypothesis and the observables collected by the sensors.
Advanced rankers such as DTDFPM~\cite{yang2016new}, employ artificial neural networks to prioritize the investigations.
Another notable example of an intelligent hypotheses ranker is the McAfee's Investigator~\cite{mcAfee}. 

\subsection{Hypotheses Testing Workflows}
\label{sec:wf}
Once an attack hypothesis has been proposed it should be tested. 
The purpose of hypothesis testing workflows is collecting forensic data, in order to complete the missing information and support or refute the hypothesis.
This is the second level of automation of the threat hunting process that reduces the human effort involved in \emph{targeted evidence acquisition}. 

Hypothesis can be confirmed by a workflow if it finds sufficient evidence that the attack described by the hypothesis takes place. 
In such a case the Ranker module (see Section~\ref{sec:ranker}) will rank the hypothesis higher than alternatives. 

It is harder to refute a hypothesis than to confirm one because the fact that we did not find sufficient evidence does not mean the respective malicious activity does not take place. 
Nevertheless, the rank of a hypothesis can be lowered if specific relevant evidence was searched for but not found.  

Workflow is a distributed algorithm that encodes the evidences acquisition logic and defines the specific investigative steps. 
Such workflows include multiple tasks that need to be executed by the agents, depending on the threat in question. 
Since workflows are (distributed) programs they also include control operators such as loops and decision points. 
Given a workflow, it may be important to distribute the tasks to different agents in order to optimize time efficiency, or balanced CPU consumption, network traffic consumption, and other resources. 
Agents may execute the various workflows in parallel, serially, or at a specific time or context as determined by the workflow execution plan.

Different scripting languages, such as Python or Java, can be used to implement workflows.  
Since workflows are automatically generated based on data (attack models and indicators) imported from external sources, they cannot be trusted. 
It is important to run automatically generated workflows within a secured container, such as RestrictedPython, in order to reduce the risk of exploits provided in form of IOCs or other CTI elements. 

Workflows contain three main types of instructions: policies, tasks and alerts.
\emph{A policy} is a set of rules and sub-rules that are part of an Intrusion Prevention System (IPS). 
Policies monitor only activities that take place after the policies' deployment.
Policies can be assigned to one or more hosts using the agents installed at every host.

\emph{A task} is an executable which is deployed upon demand and is being executed on chosen hosts in a scheduled manner.
Tasks are used to search for forensic artifacts, i.e., evidences to activities that took place in the past.

\emph{Alerts and alert handlers} are search queries that are generated to search in the SIEM for the data that the policies and tasks collect.
Alerts are being executed automatically in a scheduled manner in the SIEM.
Whenever a search query returns non empty results an alert handler receives these results 
and automatically activates the next step of the investigation as encoded in the workflow.
Namely, alert handlers are callback functions responsible for the automated transition from one stage of the investigation to the other.


\subsection{Workflow Generator}
\label{sec:w-gen}
Workflows can be hand-crafted on a case-by-case manner by a security analyst, based on relevant up-to-date CTI. 
Since such workflows are tailored to specific investigated attacks, the sophistication level required from an analyst to write them is relatively low.   
However, the large number of descriptions of potential attacks makes this approach infeasible. 

Workflows can also represent generic investigation procedures suitable for large classes of attacks similar to the approach taken by Demisto~\cite{demisto}.  
Such workflows need to be developed by skillful personnel. 
Generic workflows may be too complex to handle efficiently and may include superfluous actions that are not required for an investigation of the specific threat being confronted.
In addition, even the most generic and sophisticated workflows may become obsolete when new techniques and procedures are introduced to the cyber-warfare.    

In order to benefit from the specificity and timeliness of case-by-case workflows and reduce the human effort, 
WG (Workflow Generation) module \emph{closes the loop} by providing the third level of automation, where good hypotheses testing workflows are automatically generated based on latest CTI. 

\subsubsection{What a good workflow is?}
A good hypothesis testing workflow should, on one hand, collect all the information related to the investigated threat, 
and on the other hand, perform it in an efficient manner without interfering with the operation of the organization.  

\paragraph{Resources}
A good hypotheses testing workflow should not consume the resources of the operation critical end hosts and servers up to the level where it is noticeable by the users. 
The workflow should mind the CPU and memory consumption of the investigative tasks performed during the hunting.  
It should also consider the computational resources required to perform the analysis on the equipment dedicated for the cyber security operations.

\paragraph{Accuracy}
A good workflow should find all artifacts generated by the attacker. 
The amount of irrelevant information collected by the workflow and presented to the analyst should be minimized. 
The investigative steps performed by a workflow should be sufficient for the Ranker to give the highest rank to the hypothesis that best describes the ongoing attack.   

\paragraph{Timeliness}
A good workflow should refute or confirm as many hypotheses as possible as fast as possible.

\subsubsection{Design considerations}

\paragraph{IOC robustness vs. indicativeness}

IOCs are observables that are indicative enough to identify malicious activity. 
Liao et al.~\cite{liao2016acing} define robustness of IOCs as the percentage of IOCs that remain unchanged in a set of reported campaigns.
Most IOCs at the bottom of the PoP, such as hash values, IP addresses, domain names, and network/host artifacts, are not robust.
These indicators are easily modified by the attacker, for example, an attacker may change IP addresses and file hashes with minimal effort.\footnote{The attacker may use anonymous proxy service like Tor in order to change the IP, while a file's hash is changed by flipping a bit in an unused resource or adding spaces.}
Even domain names can easily be changed due to the free DNS providers and lax registration standards.
As a consequence, using only IOCs from the bottom of the PoP to search for a malicious activity within an environment results in high precision and low recall.

Some observables may be used to alert about a suspicious activity within the system but they cannot be considered a strong indication of a compromise.  
Such alerts are more robust matching a larger number of incidents but their precision is low due to non negligible false positive rate. 
Alert correlation and filtering techniques~\cite{saikia2019manadac,granadillo2016new,Sen2006} are commonly used to reduce the false positive rate eventually resulting in detectors with a good trade-off between precision and recall.     

Similarly, correlation between a number of robust observables, each one of which is not indicative enough to identify a compromise, may be sufficient to support an attack hypothesis.

\paragraph{CTI based automation}

While investigating an incident the analyst provides inputs to the machinery that allow gathering relevant information in order to report the summary back to the human~\cite{thomas2017dynamic,mcAfee}.
One way to reduce the involvement of a human analyst on early stages of the hypothesis testing process is by relying on CTI. 
A WG module should automatically infer from the CTI and the artifacts found so far, which information need to be collected if and when an alert is raised. 
This information is used to perform the next investigative step. 

Although, a human analyst must curate the automatic investigation process, his involvement can be gradually reduced using intelligent workflow generation. 
This will allow the analyst to focus on strategic aspects of the investigation process rather than instructing the machinery which information should be collected in response to each alert.  
The workflows generated based on the CTI should eventually collect and provide the analyst all artifacts relevant to the performed investigation.    
This process may be long and contain multiple steps and branches depending on the intermediate results of the automated investigation. 


The WG algorithms should generate workflows that perform fully automated investigations of simple attack hypotheses.

\paragraph{Performance optimization}

Consider three kinds of monitoring and data collection activities: real-time measurements, on access monitoring, and forensic data collection. 

\emph{On-access monitoring} is a kind of real-time monitoring where the IDS scans a resource, e.g. file or registry key, when this resource is used. 
Usually this kind of monitoring is implemented by hooking the system at the lowest levels. 
Since the scan is performed only when the resource is accessed and the check vs. the IOC lists is very fast, such method is considered to be least resource intensive. 
The downside of on-access monitoring is the inability to find IOCs generated before the beginning of the monitoring process. 

\emph{Real-time measurements} typically include various performance metrics such as CPU, memory, I/O, network measurements of sorts, etc. 
For example, anomaly detection algorithms employed by IDS may use these measurements to detect anomalous behavior of hosts and processes~\cite{7167219,6332015}. 
The resources consumed by the real-time monitors themselves mostly depend on the sampling rate and the method used to collect the data. 

\emph{Forensic data collection} is the most resource intensive kind of data collection activity. 
For example, in order to find whether or not a file exists on the system, we need to scan the hard-drive (or the file index if it exists). 
We include under this category the most advanced analytics and forensic investigation tools such as EnCase\footnote{https://www.guidancesoftware.com/encase-forensic} and Volatility\footnote{https://www.volatilityfoundation.org/}.

In order to save the computational resources the generated workflows should employ on-access monitoring and real time measurements (with low sampling) while looking for the first leads. 
When a lead is found the more resource intensive monitoring can be performed in a targeted manner to find the most indicative artifacts, either forensic or real-time. 
Note that ATHAFI also collects performance measurements related to the activities of its own workflows in order to avoid interfering with the operation of the organization.  






\subsection{ATHAFI for Managed Security Service Providers (MSSP)}
Organization experience increasing difficulties in hiring and maintaining skillful cyber security experts.
To address the skills gap, many organizations turn to managed security service providers (MSSPs) for outside security help~\cite{skills-gap}.
ATHAFI framework, as described so far, is designed for a single location deployment framework.
However, with minor modifications to the architecture, ATHAFI framework can be used by MSSPs (see Figure~\ref{fig:athafi4mssp}).
In such a case, the global components at ATHAFI's data center include the AttackDB, the workflows' database, global SIEM, global C\&C, and the AHG, WG and Ranker modules.
The client's on-premises components include a local SIEM that stores data and sends in batches to the global SIEM, a workflow container managed by the global C\&C, and the agents' infrastructure for data collection and incidence response.

\begin{figure}[h]
    \centering
    \includegraphics[width=0.5\textwidth]{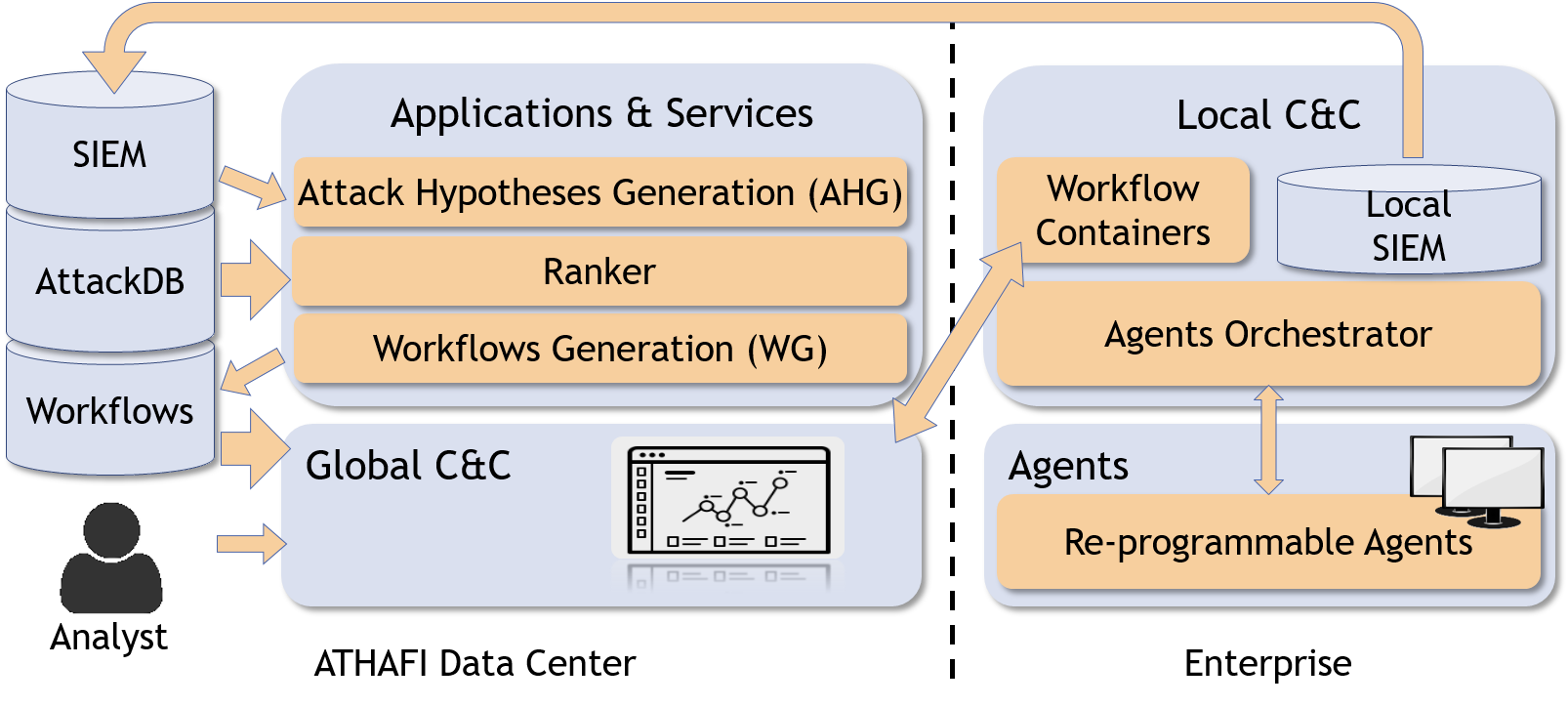} 
    \caption{ATHAFI high-level architecture for MSSP\label{fig:athafi4mssp}}
\end{figure}

\comment{
\begin{figure*}[h]
    \centering
    \subfloat[ATHAFI single location deployment (enterprise)]
    {
        \includegraphics[width=0.35\textwidth]{figures/athafi4enterprise.png} 
	    \label{fig:framework}
    }
    \subfloat[ATHAFI MSSP deployment]
    {
        \includegraphics[width=0.5\textwidth]{figures/athafi4mssp.png} 
	    \label{fig:athafi4mssp}
    }
    \caption{ATHAFI high-level architecture}
\end{figure*}
}

\section{Example Implementation and Deployment}
\label{sec:implementation}
In this section we highlight the implementation caveats of the ATHAFI functional core as well as the functional shell modules that support it.

\subsection{Building the AttackDB}
In current ATHAFI implementation, AttackDB knowledge graph is stored in a Neo4j database.
Any graph database may used for this purpose. 
We constructed a rich AttackDB that is combined of CTI from 
MITRE ATT\&CK Enterprise knowledge base\footnote{\url{https://github.com/swimlane/pyattck}}, 
Alien Vault Open Threat Exchange\footnote{\url{https://otx.alienvault.com/dashboard/new}}, 
and Virus Total\footnote{\url{https://www.virustotal.com/gui/home/search}}.

\emph{MITRE's ATT\&CK} is a CTI open knowledge base that contains information on threat adversarial techniques and tactics, threat actors, mitigation, malware, and tools~\cite{strom2017finding}.  
First we populate the AttackDB with malware, techniques, and the relationships between them extracted from MITRE ATT\&CK.

\emph{AlienVault OTX} contains CTI in the form of pulses, which contain one or more IOCs, such as file hashes, URLs, IPs. 
Pulses can be tagged with malware names, threat actors, and additional information.  
As a second step, we search for pulses using malware names via OTX API  
and link malware nodes in the AttackDB to IOCs from the respective pulses. 
The most important IOCs required for the last, third, step of populating the AttackDB are hashes. 
In current implementation we do not consider the MITRE ATT\&CK IDs specified in AlienVault pulses.

AlienVault allows anyone to post a set of IOCs as a pulse, which causes a reliability problem. 
Therefore, we used pulses uploaded by the top 20 publishers with the most subscriptions who posted pulses related to the searched malware:
AlienVault, MalwarePatrol, jnazario, niddel, Metadefender, cyberprotect, popularmalware, Malwaremustdie, Cyber\_Hat, burberry, bartblaze, ESET-Spain, julsec, zer0daydan, rpsanch, erik, milind, BLUELIV, techhelplist, BotnetExposer, and nightingale.

To further enrich the AttackDB with various network and host artifacts  observed in relation to each of the malware.  
For this purpose, in the third step, we retrieved behavioral analysis data from \emph{VirusTotal} for all hashes obtained from AlienVault.
The observed behavioral data retrieved from VirusTotal includes file names (opened, created, searched, etc.), URLs, domains, IPs, process names, registry keys, mutexes, and emails, and more.


The resulting AttackDB contains 214 malware nodes associated with IOCs, 93 malware nodes not associated with IOCs.

\begin{figure}[h]
  \centering
  \includegraphics[width=\columnwidth]{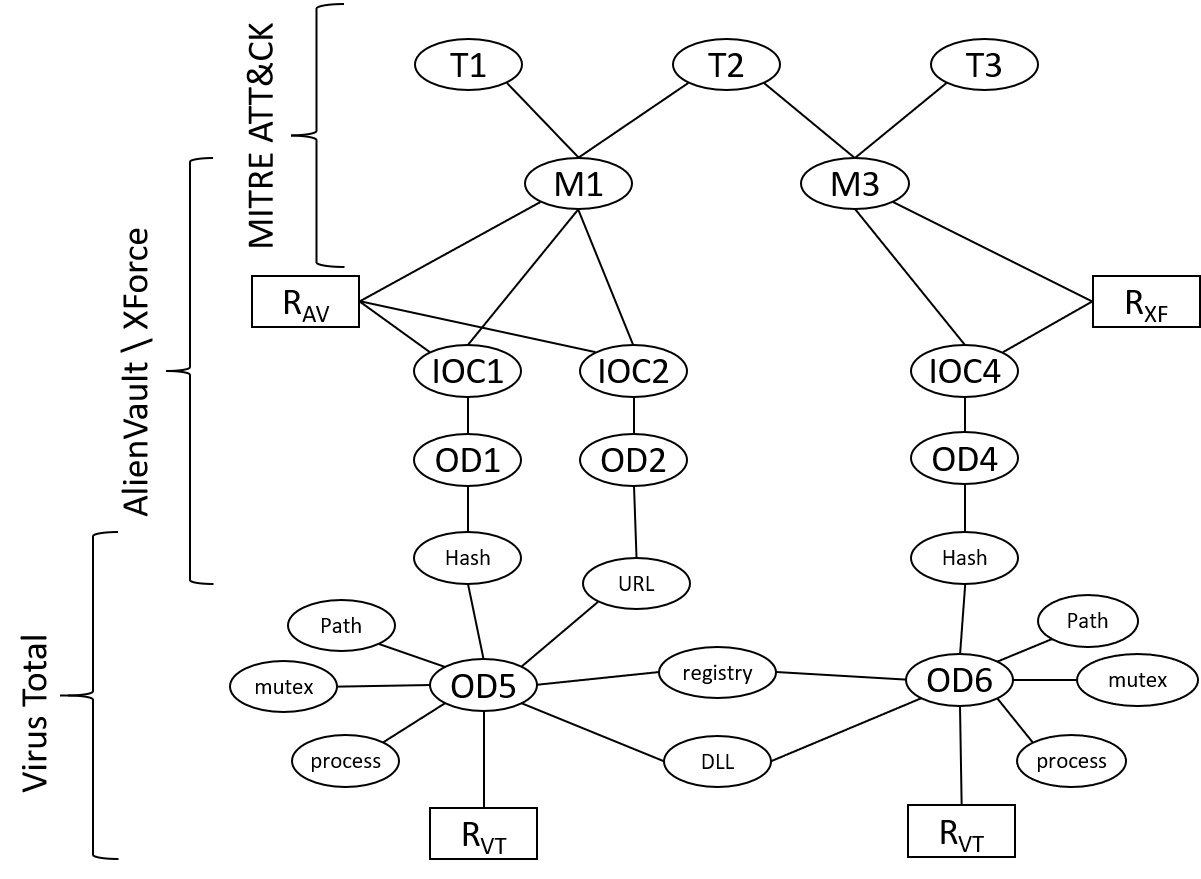}
  \caption{An illustration of the AttackDB structure}
  \label{fig:attackdb_example}
\end{figure}

Figure~\ref{fig:attackdb_example} depicts an example of the AttackDB containing two malware nodes and all the relevant connections.  
At the top level we see malware and associated techniques extracted from MITRE ATT\&CK. 
The three respective IOCs (two hash values and one URL) are extracted from AlienVault.
Following STIX format the relevant Observed Data nodes are tagged with IOCs that indicate the relevant malware. 
Finally, the behavioral data extracted from VirusTotal is displayed in the bottom of the figure. 
This observed data is not tagged with IOC nodes because these are not known as strong indications of the attack, but are merely artifacts generated by the malware during dynamic analysis. 

Note that a behavioral artifact may be connected to a malware through multiple paths. 
This happens when there are multiple instances of the same malware analyzed by VirusTotal. 
Furthermore, note that observables are indirectly connected to techniques, through the respective malware. 
The more paths are there from an observable to a technique the higher is the affinity between them. 
We will use this heuristic to build attack hypotheses in Section~\ref{sec:ahg_impl}.


\subsection{Commercial of-the-shelf components}
As a general design consideration of ATHAFI we prefer using existing tools and frameworks developed by market leaders rather than developing tailor-made in-house software for software agents' management, data collection infrastructure, and security information and event management (SIEM). 
Specifically, in this research we rely on McAfee ePolicy Orchestrator (McAfee ePO) as the agents' infrastructure and Splunk Enterprise as the SIEM.
McAfee ePO provides software agents resilient to anti-forensic malware, optimized performance, set of monitoring tools, etc. 
We chose ePO due to (1) its support of dynamic deployment of tasks on the software agents, (2) its support of deployment of specific tasks on specific agents, (3) its support of deployment of customized tasks on the agents, (4) its supply of variety of products to employ as tasks, and (5) the built-in security assurance.

Splunk Enterprise provides big-data management capabilities as well as a set of correlation, aggregation, and analysis rules. 
In ATHAFI, Splunk SIEM is responsible for receiving data created by tasks and collected according to policies executed by the ePO agents on the hosts. 
Data generated as a result of McAfee's tasks and policies is first sent to the ePO server. 
Then Splunk collects the data from the ePO server.
In addition, Splunk provides (1) an API for the automatic execution of search queries over the collected data, (2) an API for the automatic definition of rules that will alert upon a specific accumulation of data (and delegation of the alert notifications to the alert handlers).




\subsection{Workflow Execution Container (WF-Container)}
\label{sec:wf-container}

\begin{figure*}[h]
	\centering
	\includegraphics[width=0.8\textwidth]{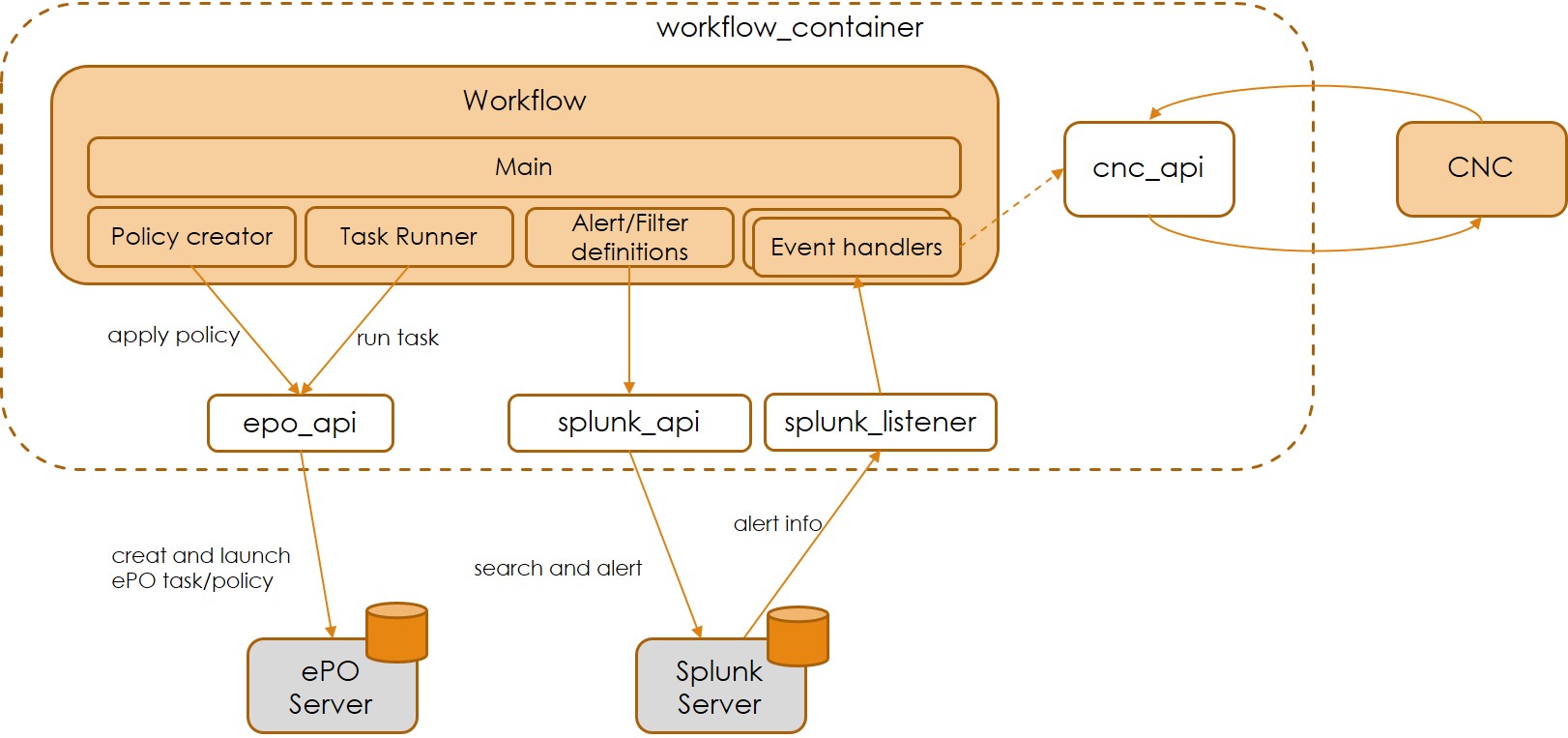}
	\caption{\label{fig:workflow-container}
		WF-Container's High-Level Architecture}
\end{figure*}

The C\&C starts a separate process of WF-Container for each workflow that should be executed.
The WF-Container receives from the C\&C a workflow and is responsible for:
(1) executing the workflow in a secure mode (RestrictedPython mode);
(2) linking the workflow commands with the ePO and Splunk servers; and
(3) returning the workflow's final execution status to the C\&C.

The WF-Container prevents the workflow from directly accessing the C\&C and the APIs provided by the ePO and Splunk software.
Instead it wraps the direct access to C\&C, ePO and Splunk using the following components (see Figure~\ref{fig:workflow-container} for illustration):
\begin{enumerate}
	\item cnc\_api – responsible for the communication with the C\&C.
	\item epo\_api – responsible for the communication with the ePO server.
	\item splunk\_api and splunk\_listener – responsible for the communication with the Splunk server.
\end{enumerate}

These components provide the following 'building blocks' for 
a workflow: running tasks and activating policies at specific hosts, defining rules for the Splunk to launch (alerts), defining handlers to alerts.

\section{Summary}
\label{sec:conc}

Threat hunting is one of the most important security operations for mitigating cyber threats in the ever expanding cyber threat landscape.
Unfortunately, many organizations do not have enough competent security analysts to perform threat hunting tasks.
Automation during threat hunting helps reducing the time and effort of the analyst, 
increasing the scale and efficiency of hunts across the enterprise, 
and reducing the required analyst qualification for performing certain types of investigations.
However, there is much automation to add in the reactive and proactive threat hunting cycle in order to significantly boost the analyst's productivity during investigation of the harshest cases.

In this paper we presented a framework for agile threat hunting and forensic investigation (ATHAFI) that adapts to the current state of the organization and state of the art threats and addresses the need for continuous forensic investigation.
This is achieved by combining high level CTI, hypotheses generation, attack hypotheses ranking, workflow generation, and targeted data collection into one unified semi-automated threat hunting process.

The proposed AttackDB schema encapsulates CTI from various sources to be used by the other modules.
The Workflow Generation (WG) module adapts the threat hunting procedures either to the latest CTI or to the likeliest attack hypotheses generated by the Attack Hypotheses Generation (AHG) module.
Thus, the AHG and the WG modules facilitate not only the identification of threat activities based on the IoAs, but also the generation of new attack hypothesis and their automatic validation. 

We define a set of requirements and APIs for the must-have functionality allowing flexible implementation of the hunting loop. 
Finally, we provide an implementation show case which integrating multiple COTS products into the unified ATHAFI framework.

	


\bibliographystyle{IEEEtran}
\bibliography{main}

\end{document}